\documentstyle[11pt]{article}

\global\arraycolsep=1pt
\begin{document}

$\frac {} {}$

\font\cmss=cmss10 \font\cmsss=cmss10 at 7pt


\hfill HUTP-96/A029

\hfill hepth/9607206

\hfill July, 1996

\begin{center}
\vspace{10pt} {\large {\bf QUANTUM TOPOLOGICAL INVARIANTS,\\[0pt]
GRAVITATIONAL INSTANTONS\\[0pt]
AND THE TOPOLOGICAL EMBEDDING }} \vspace{10pt}

{\sl Damiano Anselmi}

\vspace{4pt}

{\it Lyman Laboratory, Harvard University, Cambridge MA 02138, U.S.A.}

\vspace{12pt}

{\bf Abstract}
\end{center}

\vspace{4pt} Certain topological invariants of the moduli space of
gravitational instantons are defined and studied. Several  
amplitudes of two
and four dimensional topological gravity are computed. A notion of  
puncture
in four dimensions, that is particularly meaningful in the class of Weyl
instantons, is introduced. The {\sl topological embedding}, a  
theoretical
framework for constructing physical amplitudes that are  
well-defined order
by order in perturbation theory around instantons, is explicitly  
applied to
the computation of the correlation functions of Dirac fermions in a
punctured gravitational background, as well as to the most general  
QED and
QCD amplitude. Various alternatives are worked out, discussed and  
compared.
The quantum background affects the propagation by generating a certain
effective ``quantum'' metric. The topological embedding could  
represent a
new chapter of quantum field theory.

\vfill\eject

\section{Introduction and motivation}

\label{intro} \setcounter{equation}{0}

Given a manifold or, in general, a field configuration, one can define
topological quantities like the Pontrjiagin number and the Euler  
number. In
quantum field theory, one mainly deals with spaces of field  
configurations,
rather than single field configurations. Consequently, it can be  
interesting
to study topological invariants of such spaces. These invariants  
were called 
{\sl quantum} in ref.\ \cite{scond}, since they involve an  
integration over
the chosen configuration space. The usual topological invariants  
were called 
{\sl classical}. The quantum topological invariants are defined in a way
that is originally suggested by topological field theory, if  
treated with
the approach of ref.\ \cite{anomali}, but that actually live quite
independently. No notion of functional integral is strictly  
necessary, so
that, from the mathematical point of view, every formula is rigorously
well-defined.

Among the spaces of field configurations, special interest has to  
be devoted
to the space of instantons, or, in general, the minima of the  
action of a
physical model. The quantum topological quantities are naturally  
associated
with ``measures'' over the moduli space and such measures can be  
physically
relevant. Following the ideas developed in ref.s \cite{me,scond}  
(collected
under the name of {\sl topological embedding}), these peculiar  
measures can
be useful to define perturbation theory around instantons, bypassing
convergence problems with the moduli space integration \cite{thooft}. In
other words, new amplitudes in quantum field theory can be  
constructed and
proven to be well-defined order by order in perturbation theory around
instantons. They receive contributions only from a specific topological
sector (the one to which the instanton belongs) and could lead to
qualitatively new physical predictions.

In this paper, I investigate these issues in topological and quantum
gravity. With the new method I recover known amplitudes in 2-D  
topological
gravity and compute new ones. Then, I define and study punctures in four
dimensions and compute their characteristic quantum topological  
invariants.
Finally, I apply the topological embedding procedure to the  
computation of
various physical (i.e.\ non-topological) correlation functions,  
illustrating
many aspects of the arguments of \cite{me,scond}. I consider first the
two-point function of a fermion in a punctured background, both in  
two and
four dimensions, and then the most general QED\ and QCD\ amplitude.  
The main
effect of the quantum background on the propagation seems to be the
generation of an effective metric, that we can call the ``quantum''  
metric
of the problem. Several different quantum backgrounds are analysed and
compared.

\section{Punctures in two dimensions}

\label{punct2} \setcounter{equation}{0}

On a two dimensional plane, let us consider metrics of the form 
\begin{equation}
e^{a}=\left( 1+\sum_{i=1}^{n}{\frac{\rho  
_{i}^{2p}}{(x-x_{i})^{2p}}}\right)
^{\frac{1}{2p}}\delta _{\mu }^{a}{\rm d}x^{\mu }\equiv D^{\frac{1}{2p}%
}\,\delta _{\mu }^{a}{\rm d}x^{\mu }\equiv {\rm  
e}^{{\frac{1}{2p}}\varphi
}\,\delta _{\mu }^{a}{\rm d}x^{\mu }.  \label{p=0}
\end{equation}
The behaviour of the metric around $x_{i}$ is 
\begin{equation}
e^{a}\simeq {\frac{\rho _{i}}{|x-x_{i}|}}{\rm d}x^{a}.  \label{sing}
\end{equation}
The points $x_{i}$ are {\sl punctures} on the plane, i.e. points  
that are
sent to infinity. The scales $\rho _{i}$ specify, in some sense, the
``size'' of the puncture. Here they are just external parameters (no
topological quantity depends on them). Later on we shall discuss  
cases in
which the $\rho _{i}$ are treated as moduli, on the same footing as the
positions $x_{i}$ of the punctures. The integer number $p$ is  
introduced for
convenience. The topological quantities are independent of $p$,  
since they
are sensitive only to the ``singularities'' (\ref{sing}). Unless  
specified,
we use $p=1$ in this section. We have 
\[
\omega =-{\frac{1}{2D}}{\rm d}x^{\mu }\varepsilon _{\mu \nu  
}\partial _{\nu
}D,\quad \quad R={\rm d}\omega ,\quad \quad {\frac{1}{2\pi }}\int_{%
\relax{\rm I\kern-.18em R}^{2}}R=n. 
\]
The singularity of the spin-connection $\omega $ is a gauge artifact.
Indeed, the curvature $R=d\omega $ is everywhere regular. This  
means that
the metrics that we are considering are physically meaningful. The  
{\sl %
classical} topological invariant ${\frac{1}{2\pi }}\int_{%
\relax{\rm
I\kern-.18em R}^{2}}R$ counts the number of punctures.

The topological field theory of gravity with the metrics  
(\ref{p=0}) (see 
\cite{twist1} for the formal construction of the theory) can be solved
explicitly using the method of ref.\ \cite{anomali}, combined with some
remarks made in ref.\ \cite{scond}. There is no need to recall the  
procedure
here, since it was extensively discussed in ref.s  
\cite{anomali,me,scond}.
One finds that the diffeomorphism ghosts $c^{a}$ are 
\begin{equation}
c^{a}=-{\frac{1}{D^{\frac{1}{2}}}}\sum_{i=1}^{n}{\frac{\rho  
_{i}^{2}{\rm d}%
x_{i}^{a}}{(x-x_{i})^{2}}},\quad \quad \hat{e}^{a}\equiv  
e^{a}+c^{a}={\frac{1%
}{D^{\frac{1}{2}}}}\left[ {\rm d}x^{a}+\sum_{i=1}^{n}{\frac{\rho  
_{i}^{2}%
{\rm d}(x-x_{i})^{a}}{(x-x_{i})^{2}}}\right] .  \label{unmod}
\end{equation}
As explained in sect.\ 2 of ref.\ \cite{anomali}, everything  
follows from
the expression of this ghost. We allow the singularity in $c^{a}$  
because we
are allowing the same behaviour for $e^{a}$. However, the solution  
to the
topological field theory makes sense, since the components of the BRST
extended curvatures are all regular. The Lorentz ghosts $c^{ab}$  
are given
by 
\begin{equation}
c^{ab}={\cal D}^{[a}c^{b]},\quad \quad \hat{\omega}^{ab}\equiv  
\varepsilon
^{ab}\hat{\omega}=\omega ^{ab}+c^{ab}=\varepsilon ^{ab}{\frac{1}{D}}%
\sum_{i=1}^{n}{\frac{\rho _{i}^{2}{\rm d}(x-x_{i})^{\mu  
}\varepsilon _{\mu
\nu }(x-x_{i})^{\nu }}{(x-x_{i})^{4}}}  \label{first}
\end{equation}
One can easily check that the first descendant of the torsion,  
called $\psi
^{a}$ in ref.\ \cite{twist1}, is indeed regular: 
\begin{equation}
\psi ^{a}=-{\cal D}^{\{a}c^{b\}}e^{b}+{\frac{1}{2}}s\varphi  
\,e^{a},\quad
\quad s=\sum_{i=1}^{n}{\rm d}x_{i}^{\mu }{\frac{\partial }{\partial
x_{i}^{\mu }}},  \label{second}
\end{equation}
$s$ being the exterior derivative on the moduli space. The BRST extended
curvature, which is the basic ingredient in the construction of the
observables, is (${\rm \hat{d}}={\rm d}+s$) 
\begin{eqnarray}
{\hat{R}} &=&{\rm \hat{d}}\hat{\omega}={\frac{2}{D^{2}}}\left[  
\sum_{i=1}^{n}%
{\frac{\rho _{i}^{2}{\rm d}^{2}(x-x_{i})}{(x-x_{i})^{4}}}\left(  
D-{\frac{%
\rho _{i}^{2}}{(x-x_{i})^{2}}}\right) +\right.  \nonumber \\
&&\left. \sum_{i\neq j}{\frac{\rho _{i}^{2}\rho _{j}^{2}}{%
(x-x_{i})^{4}(x-x_{i})^{4}}}(x-x_{i})\cdot {\rm d}(x-x_{i})\,\,{\rm d}%
(x-x_{j})^{\mu }\varepsilon _{\mu \nu }(x-x_{j})^{\nu }\right] .
\label{curva}
\end{eqnarray}
The observables are 
\[
{\cal O}_{\gamma }^{(d)}={\frac{1}{(2\pi )^{d}}}\int_{\gamma  
}{\hat{R}}^{d}, 
\]
where $\gamma $ can also be a point, in which case the observable  
is denoted
by ${\cal O}^{(d)}(x)$ (simply ${\cal O}(x)$ for $d=1$).

The correlator $<\prod_{j=1}^{n}{\cal  
O}(y_{j})>={\frac{1}{n!}}{\cal A}_{n}$
with local observables placed in distinct points is equal to one.  
This can
be proved as follows. Let us write ${\cal O}(y_{n})$ as ${\frac{1}{2}}%
\varepsilon _{ab}sc^{ab}(y_{n})$ and integrate by parts on the  
moduli space.
The singularities of $c^{ab}$ in $x_{i}$ are responsible for the
nonvanishing contributions. Each puncture contributes the same, so  
that we
have $n$ times a reduced amplitude ${\frac{1}{n!}}{\cal A}_{n-1}$  
in which
one modulus, say $x_{n}$, disappears. In ${\cal A}_{n-1}$ one has  
to set $%
dx_{n}=0$ and take the limit $y_{n}\rightarrow x_{n}$. Before  
taking this
limit one has to note that ${\frac{1}{(n-1)!}}{\cal A}_{n-1}$ is  
the same as 
$<\prod_{j=1}^{n-1}{\cal O}(y_{j})>$ with the replacements $\rho
_{i}^{2}\rightarrow {\frac{\rho _{i}^{2}}{1+\rho  
_{n}^{2}/(y_{n}-x_{n})^{2}}}
$, $i=1,\ldots n-1$. This can be proven by direct inspection of  
(\ref{curva}%
). After repeating the argument $n-1$ times, one arrives at an  
expression $%
{\cal A}_{1}$, which is the same as $<{\cal O}(y_{1})>$ with $\rho
_{1}^{2}\rightarrow {\frac{\rho _{1}^{2}}{\prod_{i=2}^{n}(1+\rho
_{i}^{2}/(y_{i}-x_{i})^{2})}}$. ${\cal A}_{1}$ is equal to one,  
whatever $%
\rho _{1}$ is. At this point the limits $y_{i}\rightarrow x_{i}$, $%
i=2,\ldots n$ are all trivial and the final result is also one, as  
claimed.

When there are observables placed in coincident points, the above  
procedure
cannot be applied and we have to do the computation in a different  
way. A
long algebraic manipulation similar to the ones described in detail  
in ref.\ 
\cite{scond}\footnotemark 
\footnotetext{
In particular, one has to use formul\ae\ (4.24) and (4.26) of that  
paper.}
gives 
\begin{equation}
{\hat{R}}^{n}={\frac{2^{n}n!}{D^{n+1}}}\prod_{i=1}^{n}{\frac{\rho  
_{i}^{2}%
{\rm d}^{2}(x-x_{i})}{(x-x_{i})^{4}}}.  \label{formu}
\end{equation}
The correlation functions $<\prod_{j=1}^{k}{\cal  
O}^{(d_{j})}(y_{j})>$ can
be computed using the cluster property. The independence of the  
points $%
y_{j} $ assures that we can take the limits  
$|y_{j}-y_{k}|\rightarrow \infty 
$ $\forall j\neq k$. This shows that 
\[
<\prod_{j=1}^{k}{\cal O}(y_{j})^{d_{j}}>=\prod_{j=1}^{k}<{\cal O}%
(y_{j})^{d_{j}}>. 
\]
Formula (\ref{formu}) gives 
\begin{equation}
<[{\cal O}^{(n)}(x)]>={\frac{1}{(2\pi )^{n}n!}}\int  
{\frac{2^{n}n!}{D^{n+1}}}%
\prod_{i=1}^{n}{\frac{\rho _{i}^{2}{\rm  
d}^{2}x_{i}}{(x-x_{i})^{4}}}={\frac{1%
}{n!}},  \label{urc}
\end{equation}
so that 
\begin{equation}
<\prod_{j=1}^{k}{\cal  
O}(y_{j})^{d_{j}}>=\prod_{j=1}^{k}{\frac{1}{d_{j}!}}.
\label{formula}
\end{equation}
These are the (known) correlation functions of 2D topological  
gravity \cite
{kontsevich}, here computed on the plane, instead of the sphere. To  
recover
the ones on the sphere, one has to remember that three ``hidden''  
punctures
are not visible in our approach, since they just do the job of  
fixing the
gauge of the ghosts. Moreover, the usual normalization is to  
multiply the
amplitude by the symmetry factor $n!$.

Before proceeding, let us make a couple of comments. The moduli  
space has
diagonal subspaces $\Delta $ where the positions of the punctures  
coincide.
Discarding $\Delta $ by saying that it is of vanishing measure  
could be too
naive, since the measure could be singular on $\Delta $ (the  
measure is a
descendant of the Euler characteritic, which jumps on $\Delta $).  
Actually,
we did not find any problem in our computation and this can be  
interpreted
as follows. The jump of the Euler characteristic is transferred, at the
quantum level, to a jump of the correlation functions, that takes  
place when
the positions of the local observables coincide. In each correlation
function $\Delta $ is of vanishing measure, but there are many different
correlation functions to consider, i.e.\ many measures. One can  
obtain some
topological information about $\Delta $ by comparing the values for
different sets of coindident points. In particular, we have learnt  
that when 
$n$ points come to coincide, there is a jump of $1/n!$. I argue  
that this is
a completely general feature of punctures (and of the centers of  
Yang-Mills
instantons, as we shall see), because the same thing happens in four
dimensions (see section \ref{punct4}). In the case of vortices, on  
the other
hand, there is no jump at all \cite{scond}. Exotic values of the  
jumps will
be also found in the next section.

Now, let us study some correlation functions with nonlocal observables.

I consider the plane with one puncture and a correlation function  
$<{\cal O}%
_{\gamma _{1}}\cdot {\cal O}_{\gamma _{2}}>$, $\gamma _{1}$ and  
$\gamma _{2}$
being two curves. The explicit evaluation gives 
\begin{equation}
<{\cal O}_{\gamma _{1}}\cdot {\cal O}_{\gamma _{2}}>={\frac{\rho  
^{4}}{\pi
^{2}}}\int_{\gamma _{1}}\int_{\gamma _{2}}\varepsilon _{\mu \nu  
}{\rm d}%
x^{\mu }{\rm d}y^{\nu }\int {\frac{{\rm d}^{2}x_{0}}{(\rho
^{2}+(x-x_{0})^{2})^{2}(\rho ^{2}+(y-x_{0})^{2})^{2}}}.  \label{2.11}
\end{equation}
To understand what this means, it can be useful to take the limit $\rho
\rightarrow 0$, where one gets 
\[
<{\cal O}_{\gamma _{1}}\cdot {\cal O}_{\gamma _{2}}>=\int_{\gamma
_{1}}\int_{\gamma _{2}}\varepsilon _{\mu \nu }{\rm d}x^{\mu }{\rm  
d}y^{\nu
}\delta (x-y). 
\]
We see that this amplitude counts the number of intersections of $\gamma
_{1} $ and $\gamma _{2}$ with signs. Roughly speaking, this is  
similar to
the link numbers of \cite{anomali,me}. Indeed, in the multilink
interpretation offered in \cite{me}, the importance of  
intersections among
the $\gamma $'s was apparent. Borrowing the graphical notation from  
\cite{me}%
, we shall denote the intersection of the two curves with  
$\backslash \!\!\!%
\slash (\gamma _{1},\gamma _{2})$. One the sphere or the plane,  
this number
is always zero, if $\gamma _{1}$ and $\gamma _{2}$ are both compact. On
higher genus Riemann surfaces it is not so. Later on we shall  
exhibit the
full set of amplitudes of this kind for any genus and and arbitrary  
(even)
number of curves. With Yang-Mills instantons \cite{anomali,me}, on  
the other
hand, there is one more modulus (the scale) and one more element in the
amplitude (the Chern-Simons form), so that it is not necessary to have a
genus in order to produce a nontrivial linkage \cite{anomali}. Something
similar will be observed below.

With more nonlocal observables, one can consider a multiple amplitude.
Straightforward manipulations allow us to show that 
\begin{equation}
<\prod_{i=1}^{2n}{\cal O}_{\gamma_i}>= \sum_{{\scriptsize {\rm %
cyclic\,\,perm \,\,of}}\, \{2,\ldots 2n\}} \backslash\!\!\!\slash %
(\gamma_1,\gamma_2)\cdots \backslash\!\!\!\slash  
(\gamma_{2n-1},\gamma_{2n}).
\label{tree}
\end{equation}
The fact that the multiple amplitude can be written purely in terms  
of the
two-amplitude suggests that $\backslash\!\!\!\slash 
(\gamma_1,\gamma_2)$ is in some sense the topological  
``propagator'' between
curves. Then, (\ref{tree}) is a ``tree-level'' amplitude. Similar  
phenomena
take place in higher dimensions.

Finally, let us analyse what happens when treating the scales $\rho  
_{i}$ as
true moduli. (\ref{unmod}) and (\ref{first}) are unchanged, while  
in (\ref
{second}) the moduli space derivative $s$ has to be replaced by 
\[
s=\sum_{i=1}^{n}{\rm d}x_{i}^{\mu }{\frac{\partial }{\partial  
x_{i}^{\mu }}}+%
{\rm d}\rho _{i}^{2}{\frac{\partial }{\partial \rho _{i}^{2}}}. 
\]
(\ref{curva}) has to be modified consequently. It is easy to check  
that the
quantum topological invariants are link numbers, similarly to the case
worked out in ref.s \cite{anomali,me}. To be concrete, let us take  
$n=1$, $%
p=1$ and the observable $<{\cal O}_{\gamma }\cdot {\cal O}(x)>$. The
computation can be done writing ${\cal O}(x)$ as  
${\frac{1}{2}}\varepsilon
_{ab}sc^{ab}$ and integrating by parts, or by direct inspection of the
integral below, which I write down explicitly for later use in  
connection
with the topological embedding. We have 
\begin{equation}
<{\cal O}_{\gamma }\cdot {\cal O}(x)>={\frac{1}{4\pi  
^{2}}}\int_{\gamma }%
{\rm d}y^{\mu }\varepsilon _{\mu \nu }(y-x)^{\nu }\int {\frac{2\rho  
^{2}\,%
{\rm d}\rho ^{2}\,{\rm d}^{2}x_{0}}{(\rho ^{2}+(y-x_{0})^{2})^{2}(\rho
^{2}+(x-x_{0})^{2})^{2}}}=\backslash \!\!\!\slash (\gamma ,\{x\}),
\label{quantumbackground}
\end{equation}
where $\backslash \!\!\!\slash (\gamma ,\{x\})={\frac{1}{2\pi  
}}\int_{\gamma
}{\rm d}y^{\mu }\varepsilon _{\mu \nu }(y-x)^{\nu }/ (x-y)^{2}$.

If one is interested in the punctured torus, one can use the metrics 
\begin{eqnarray}
e^{1} &=&({\rm d}\xi +\tau _{1}\,{\rm d}\eta )\,{\rm e}^{{\frac{1}{2p}}%
\varphi },\quad \quad e^{2}=\tau _{2}\,{\rm d}\eta \,{\rm  
e}^{{\frac{1}{2p}}%
\varphi },  \nonumber \\
\varphi &=&\ln \left( 1+\sum_{i=1}^{n}\rho  
_{i}^{2p}F_{p}(x-x_{i})\right) ,
\label{toro} \\
F_{p}(\lambda ) &=&\sum_{k,l}{\frac{1}{(\lambda +k\hat{1}+l\tau  
)^{2p}}}. 
\nonumber
\end{eqnarray}
Here $\xi ,\eta \in [0,1]$, $x=(\xi +\tau _{1}\eta ,\tau _{2}\eta  
)$, $\hat{1%
}=(1,0)$, $\tau =(\tau _{1},\tau _{2})$, $x_{i}=(\xi _{i}+\tau _{1}\eta
_{i},\tau _{2}\eta _{i})$, $\xi _{i},\eta _{i}\in [0,1]$. $\tau $ is the
modulus of the torus, $(\xi _{i},\eta _{i})$ are the positions of the
punctures. The convergence of the sum in $F_{p}$, in which $k$ and  
$l$ range
over the integers, requires now $p>1$. In the usual conventions,  
the above
metric describes the torus with $n+1$ punctures. The extra  
puncture, that
fixes the translations, is not visible in our approach.

The case $n=0$ has been discussed in sect.\ of \cite{anomali}, the local
observable being the Poincar\`{e} metric 
\[
{\rm Pm}={\frac{{\rm d}^{2}\tau }{(\tau -\bar{\tau})^{2}}} 
\]
and the amplitude being the volume of the moduli space (equal to  
$1/24$). Pm
is a good observable in presence of any number of punctures. It  
corresponds
to the top Chern class of the Hodge bundle and can be brought into  
the game 
{\sl via} a mechanism described in \cite{constr}. The local  
observable is,
instead equal to the Poincar\'{e} metric plus a certain remnant.

Explicit computations are much more involved, now. A simple case is  
given by
the amplitudes 
\[
\backslash \!\!\!\slash (\gamma _{1},\ldots ,\gamma _{2n})=<{\rm  
Pm}\cdot
\prod_{j=1}^{2n}{\cal O}_{\gamma _{j}}>. 
\]
that are equal to (\ref{tree}) times a numerical factor, now on the  
torus
instead of the sphere. Similar considerations easily extend to  
higher genera.

\section{Punctures in four dimensions}

\label{punct4} \setcounter{equation}{0}

In four dimensions, on $\relax{\rm I\kern-.18em R}^{4}$, we consider the
conformally flat metrics 
\begin{equation}
e^{a}=\left( 1+\sum_{i=1}^{n}{\frac{\rho  
_{i}^{2p}}{(x-x_{i})^{2p}}}\right)
^{\frac{1}{p}}\delta _{\mu }^{a}{\rm d}x^{\mu }\equiv {\rm  
e}^{{\frac{1}{p}}%
\varphi }\,\delta _{\mu }^{a}{\rm d}x^{\mu }.  \label{p=1}
\end{equation}
For any $p$ these metrics are in the same topological class (their
topological invariants are the same). The Einstein action is finite  
(zero in
the special case $p=1$), the Weyl action is zero. From the  
computative point
of view, $p=2$ will be the most convenient choice. The metrics with  
$p=1$
are discussed in ref.\ \cite{GibPop} from a classical point of view.

The singularity that we allow around a puncture $x_{i}$ is now 
\begin{equation}
e^{a}\simeq {\frac{\rho _{i}^{2}}{(x-x_{i})^{2}}}{\rm d}x^{a}.   
\label{si}
\end{equation}
While the two-dimensional puncture (\ref{sing}) describes a  
cylinder, as one
can see after the change of coordinates $|x-x_{i}|\rightarrow {\rm e}%
^{-|x-x_{i}|}$, the four-dimensional punture (\ref{si}) corresponds  
to an $%
\relax{\rm I\kern-.18em R}^{4}$, as it is clearly visible after the
inversion $(x-x_{i})_{\mu }\rightarrow (x-x_{i})_{\mu }\rho
_{i}^{2}/(x-x_{i})^{2}$.

The definition (\ref{p=1}) will be justified by the results that we  
shall
obtain. Moreover, it is such that the statement ``$\relax{\rm  
I\kern-.18em R}%
^{4}$ is $S^{4}$ with one puncture at infinity'' is literally correct.
Indeed, let us write the metric of $S^{4}$ as ${\rm d}s^{2}={\rm d}%
x^{2}/(1+x^{2}/R^{2})^{2}$. We perform an inversion $x^{\mu  
}\rightarrow {%
\frac{R^{2}}{x^{2}}}x^{\mu }$ in order to exchange the point at infinity
with zero. Then we place a puncture in the origin, multiplying the  
metric by
a conformal factor of the type (\ref{p=1}) with $p=1$, precisely  
$\left( 1+{%
\frac{R^{2}}{x^{2}}}\right) ^{2}$. Finally, we undo the inversion.  
At the
end we get the metric ${\rm d}s^{2}={\rm d}x^{2}$ of $\relax{\rm I\kern%
-.18emR}^{4}$, as desired.

A conformal factor like the one in (\ref{p=1}) ``puncturizes'' a Weyl
instanton while remaining in the class of Weyl instantons. The  
process of
``puncturization'' encoded in (\ref{p=1}) is completely general. Given a
manifold $M$, we denote by p$M$ the punctured version of the same.  
When we
want to specify the number of punctures, we write p$M_{n}$. If the
topological invariants of $M$ are $(\chi ,\sigma )$, then those of  
p$M_{n}$
are $(\chi -2n,\sigma )$. For example, interesting Weyl instantons with
punctures are the pALE (punctured Asymptotically Locally Euclidean)
manifolds, p$K3$, p$T^{4}$, p$CP^{2}$, p$S^{4}$, etc. The topological
properties that will be derived for p$\relax{\rm I\kern-.18em  
R}^{4}$ are
quite general and hold for any punctured manifold p$M$.

The hyperK\"ahler character of a gravitational instanton, instead,  
in {\sl %
not} preserved by the puncturization process. I have not found, yet, a
definition of puncture that does this job. It would be interesting  
to know
if it exists.

Coming back to p$\relax{\rm I\kern-.18em R}^{4}$, we have that the spin
connection $\omega ^{ab}$ is 
\[
\omega ^{ab}=-{\frac{1}{p}}{\rm d}x^{a}\partial _{b}\varphi  
+{\frac{1}{p}}%
{\rm d}x^{b}\partial _{a}\varphi . 
\]
The curvature $R^{ab}=d\omega ^{ab}-\omega ^{ac}\omega ^{cb}$ is  
regular.
The Pontrijagin number vanishes, while the Euler number counts the  
number of
punctures\footnotemark 
\footnotetext{
In the standard normalization, $\chi=-2n$ and $\sigma=0$. However,  
from the
point of view of topological field theory it is convenient to  
normalize the
invariants like in (\ref{norm}).}: 
\begin{equation}
-{\frac{1}{64\pi ^{2}}}\int_{\relax{\rm I\kern-.18em R}^{4}}R^{ab}\wedge
R^{cd}\varepsilon _{abcd}=n,\quad \quad \int_{\relax{\rm  
I\kern-.18em R}%
^{4}}R^{ab}\wedge R^{ab}=0.  \label{norm}
\end{equation}

Now, let us solve the topological field theory of gravity  
\cite{twist1} with
the metrics (\ref{p=1}), following the general recipe of ref.s \cite
{anomali,scond}. The ghosts $c^{a}$ of diffeomorphisms are 
\begin{equation}
c^{a}=-{\frac{1}{D^{\frac{p-1}{p}}}}\sum_{i=1}^{n}{\frac{\rho  
_{i}^{2p}{\rm d%
}x_{i}^{a}}{(x-x_{i})^{2p}}},\quad \quad \hat{e}^{a}\equiv  
e^{a}+c^{a}={%
\frac{1}{D^{\frac{p-1}{p}}}}\left[ {\rm  
d}x^{a}+\sum_{i=1}^{n}{\frac{\rho
_{i}^{2p}{\rm d}(x-x_{i})^{a}}{(x-x_{i})^{2p}}}\right] .  \label{unmod2}
\end{equation}
The expressions of the other fields\ greatly simplify for $p=2$,  
which is
the value that we shall use from now on. The Lorentz ghosts $c^{ab}$ are
again given by the first expression of formula (\ref{first}), so that 
\[
\hat{\omega}^{ab}=-{\frac{2}{D}}\sum_{i=1}^{n}{\frac{\rho _{i}^{4}}{%
(x-x_{i})^{6}}}[(x-x_{i})^{a}{\rm d}(x-x_{i})^{b}-(x-x_{i})^{b}{\rm d}%
(x-x_{i})^{a}]. 
\]
The first BRST descendant of the torsion $\psi ^{a}$, again given  
by (\ref
{second}), is regular. Using the notation $\hat{x}_{i}=x-x_{i}$ in  
order to
compress the formula, the BRST extended curvature  
${\hat{R}}^{ab}=\hat{d}%
\hat{\omega}^{ab}-\hat{\omega}^{ac}\hat{\omega}^{cb}$ is given by 
\begin{eqnarray*}
{\hat{R}}^{ab} &=&-{\frac{4}{D^{2}}}\left[  
\sum_{i=1}^{n}{\frac{\rho _{i}^{4}%
}{\hat{x}_{i}^{6}}}\left( D-{\frac{\rho  
_{i}^{4}}{\hat{x}_{i}^{4}}}\right)
\left( {\rm d}\hat{x}_{i}^{a}{\rm  
d}\hat{x}_{i}^{b}-6{\frac{\hat{x}_{i}\cdot 
{\rm d}\hat{x}_{i}}{\hat{x}_{i}^{2}}}\hat{x}_{i}^{[a}{\rm d}\hat{x}%
_{i}^{b]}\right) \right. \\
&&\left. +\sum_{i\neq j}{\frac{\rho _{i}^{4}\rho  
_{j}^{4}}{\hat{x}_{i}^{6}%
\hat{x}_{j}^{6}}}\left( 4\hat{x}_{i}\cdot {\rm d}\hat{x}_{i}\,\hat{x}%
_{j}^{[a}{\rm  
d}\hat{x}_{j}^{b]}-\hat{x}_{i}^{a}\hat{x}_{j}^{b}\,{\rm d}\hat{%
x}_{i}\cdot {\rm d}\hat{x}_{j}-\hat{x}_{i}\cdot \hat{x}_{j}\,{\rm  
d}\hat{x}%
_{i}^{a}{\rm d}\hat{x}_{j}^{b}-2\hat{x}_{j}\cdot {\rm  
d}\hat{x}_{i}\,\hat{x}%
_{i}^{[a}{\rm d}\hat{x}_{j}^{b]}\right) \right] .
\end{eqnarray*}
The observables are 
\[
{\cal O}_{\gamma }=-{\frac{1}{64\pi ^{2}}}\int_{\gamma  
}\hat{R}^{ab}\wedge {%
\hat{R}}^{cd}\varepsilon _{abcd}\equiv \int_{\gamma }\hat{{\cal Q}}(x). 
\]

The correlator $<\prod_{j=1}^{n}{\cal O}(y_{j})>$, $y_{j}\neq  
y_{k}$ for $%
j\neq k$, is equal to one. This can be proved exactly like in the
two-dimensional case, by a sequence of partial integrations and  
rescalings
of the $\rho _{i}$'s. The detailed derivation is left to the  
reader. When
there are coincident points, there should exist a simple formula
generalizing (\ref{formu}), but it seems very difficult to work it  
out. In
the case $n=2$, a long work done with Mathematica gives 
\[
\left[ \hat{{\cal Q}}(x)\right] ^{2}={\frac{384}{\pi ^{4}}}{\frac{\rho
_{1}^{8}\rho  
_{2}^{8}d^{4}\hat{x}_{1}d^{4}\hat{x}_{2}}{D^{7}\hat{x}_{1}^{12}%
\hat{x}_{2}^{12}}}\left( D+{\frac{\rho _{1}^{4}\rho _{2}^{4}}{2\hat{x}%
_{1}^{4}\hat{x}_{2}^{4}}}-(\hat{x}_{1}\cdot \hat{x}_{2})^{2}{\frac{\rho
_{1}^{4}\rho _{2}^{4}}{2\hat{x}_{1}^{6}\hat{x}_{2}^{6}}}\right) . 
\]
This allows us to compute the correlation function with two local
observables placed in the same point, which turns out to be  
(putting $\rho
_{1}=\rho _{2}=1$) 
\begin{equation}
<[{\cal O}(x)]^{2}>={\frac{1}{2!}}\int {\frac{384}{\pi  
^{4}}}{\frac{{\rm d}%
^{4}x_{1}{\rm  
d}^{4}x_{2}}{D^{7}\hat{x}_{1}^{12}\hat{x}_{2}^{12}}}\left( D+{%
\frac{1}{2\hat{x}_{1}^{4}\hat{x}_{2}^{4}}}-{\frac{(\hat{x}_{1}\cdot  
\hat{x}%
_{2})^{2}}{2\hat{x}_{1}^{6}\hat{x}_{2}^{6}}}\right) ={\frac{1}{2}}.
\label{jump}
\end{equation}
We see that the kind of jump that characterizes punctures in four  
dimensions
is the same as in two, as we wanted to show. Very presumably,  
formula (\ref
{formula}) also holds.

The Pontrijagin number of our metrics vanishes. However, this is not
sufficient to say that all the related quantum topological observables 
\begin{equation}
\tilde{{\cal O}}_{\gamma }^{(d)}=-{\frac{1}{(32\pi  
^{2})^{d}}}\int_{\gamma }%
{\rm tr}[{\hat{R}}^{2d}]  \label{tilde}
\end{equation}
give vanishing correlation functions. $<\tilde{{\cal O}}^{(1)}(x)>$
trivially vanishes. One can explicitly check that $<\tilde{{\cal  
O}}^{(1)}(x)%
\tilde{{\cal O}}^{(1)}(y)>$ also vanishes. When $x\neq y$, this is
straightforward. When $x=y$ similar algebraic manipulations as  
above give an
expression proportional to 
\begin{equation}
\int {\frac{{\rm d}^{4}x_{1}{\rm  
d}^{4}x_{2}}{D^{7}\hat{x}_{1}^{12}\hat{x}%
_{2}^{12}}}\left(  
D-{\frac{2}{\hat{x}_{1}^{4}\hat{x}_{2}^{4}}}+{\frac{2(\hat{%
x}_{1}\cdot  
\hat{x}_{2})^{2}}{\hat{x}_{1}^{6}\hat{x}_{2}^{6}}}\right) =0.
\label{ult}
\end{equation}
However, due to the identity $\left[ {\cal O}^{(1)}(x)\right]  
^{2}=2\left[ 
\tilde{{\cal O}}^{(1)}(x)\right] ^{2}+4\tilde{{\cal O}}^{(2)}(x)$,  
$\tilde{%
{\cal O}}^{(2)}(x)$ is related to a couple of ${\cal O}^{(1)}(x)$ at
coincident points, so that we have 
\begin{equation}
<\tilde{{\cal O}}^{(2)}(x)>={\frac{1}{8}}.  \label{ultima}
\end{equation}
This means that Pontrjiagin and Euler numbers mix at the quantum level,
although they are classically distinguished. This mixing does not  
violate
parity conservation, since it takes place only with even powers.

It can also be checked that ${\cal O}(x)\tilde{{\cal O}}(x)\equiv 0$. In
particular, decomposing the Riemann curvature $R^{ab}$ into the  
self-dual
and anti-self-dual components, $R^{ab}=R^{+ab}+R^{-ab}$, one can  
write $<%
{\cal O}(x)\tilde{{\cal O}}(x)>\propto <{\rm  
tr}[({\hat{R}}^{+})^{2}]^{2}-%
{\rm tr}[({\hat{R}}^{-})^{2}]^{2}>=0$. This equality, toghether  
with, (\ref
{jump}) and (\ref{ult}), implies that ${\frac{1}{(16\pi  
^{2})^{2}}}<{\rm tr}%
[({\hat{R}}^{\pm })^{2}]^{2}>={\frac{1}{2}}$. We point out this  
fact for the
following reason. There is a deep relationship between punctures and the
centers of Yang-Mills instantons. Let us consider (\ref{p=1}) for  
$p=1$. It
can be easily checked that $R^{+ab}$ turns out to coincide with the  
field
strength of the 't~Hooft $SU(2)$ Yang-Mills instantons with  
instanton number 
$n$ \cite{rebbi}, while $R^{-ab}$ is the ``conjugate'' of the same,  
namely
the `t~Hooft solution with instanton number $-n$. Instantons and
anti-instantons are placed in the same points, so that we can say that a
puncture can be viewed as an instanton-anti-instanton pair\footnotemark 
\footnotetext{
This explains also why $\chi=-2n$: a contribution $-n$ comes from  
instantons
and another contribution $-n$ comes from anti-instantons. These two
contributions cancel in the Pontrjiagin number: $\sigma=-n+n=0$.}.  
We have
just shown that ${\frac{1}{(16\pi ^{2})^{2}}}<{\rm tr}[({\hat{R}}^{\pm
})^{2}]^{2}(x)>={\frac{1}{2}}$, while ${\frac{1}{(16\pi  
^{2})^{2}}}<{\rm tr}%
[({\hat{R}}^{\pm })^{2}](x)\,{\rm tr}[({\hat{R}}^{\pm  
})^{2}](y)>=1$, for $%
x\neq y$. This result has been obtained for $p=2$, but it is  
independent of $%
p$, as we know. We conclude that the elementary jump that characterizes
punctures characterizes the centers of Yang-Mills instantons also.

Similarly to the two-dimensional case, we can consider amplitudes  
associated
with non-local observables. For example, in the case with  
one-puncture, we
get 
\[
<{\cal O}_{\gamma _{3}}\cdot {\cal O}_{\gamma _{1}}>={\frac{24}{\pi  
^{4}}}%
\int_{\gamma _{3}}\int_{\gamma _{1}}\varepsilon _{\mu \nu \rho  
\sigma }{\rm d%
}x^{\mu }{\rm d}x^{\nu }{\rm d}x^{\rho }{\rm d}y^{\sigma }\int  
{\frac{\rho
^{16}(x-x_{0})^{4}(y-x_{0})^{4}\,{\rm d}^{4}x_{0}}{(\rho
^{4}+(x-x_{0})^{4})^{4}(\rho ^{4}+(y-x_{0})^{4})^{4}}}, 
\]
$\gamma _{3}$ and $\gamma _{1}$ here denoting three- and one-dimensional
closed submanifolds (eventually not compact). A clearer  
representation of
the above correlation function is obtained by taking the limit $\rho
\rightarrow 0$ and using the property $\lim_{\rho \rightarrow  
0}{\frac{\rho
^{8}x^{4}}{(\rho ^{4}+x^{4})^{4}}}={\frac{\pi ^{2}}{2}}\,\delta  
^{(4)}(x)$: 
\[
<{\cal O}_{\gamma _{3}}\cdot {\cal O}_{\gamma _{1}}>={\frac{1}{6}}%
\int_{\gamma _{3}}\int_{\gamma _{1}}\varepsilon _{\mu \nu \rho  
\sigma }{\rm d%
}x^{\mu }{\rm d}x^{\nu }{\rm d}x^{\rho }{\rm d}y^{\sigma }\,\delta
(x-y)=\backslash \!\!\!\slash (\gamma _{3},\gamma _{1}). 
\]
So, there is a topological propagator between pairs of closed  
submanifolds
whose dimensions sum up to four. The above correlation function is more
meaningful on a topologically nontrivial manifold, like $K3$,  
$T^{4}$, $%
CP^{2}$. Using a procedure similar to the one used in section  
\ref{punct2}
for $T^{2}$, we can produce the above correlation function on any  
punctured
manifold p$M$, after inserting an additional top-form of the moduli  
space of 
$M$. On $T^{4}$ formul\ae\ generalizing (\ref{toro}) can be easily  
written
down.

When the $\rho _{i}$ are included in the set of the moduli, the quantum
topological invariants are link numbers, again. With one puncture, one
recovers exactly the multilink correlation functions of  
\cite{anomali,me}.
For example, 
\[
<{\cal O}_{\gamma _{3}}\cdot {\cal O}(x)>=\backslash \!\!\!\slash  
(\gamma
_{3},\{x\}). 
\]
With more punctures, the problem is equivalent to the same problem with
't~Hooft multi-instantons \cite{rebbi}, i.e.\ $SU(2)$ Yang-Mills  
instantons
in which one keeps the gauge moduli fixed and studies only positions and
scales.

Before closing this section, I would like to comment about the non
uniqueness of the four dimensional puntures (\ref{si}), and the  
uniqueness
of the two dimentional punctures (\ref{p=0}). To this purpose, let  
us modify
(\ref{p=1}) into 
\begin{equation}
e^{a}=\left( 1+\sum_{i=1}^{n}{\frac{\rho  
_{i}^{2p}}{(x-x_{i})^{2p}}}\right)
^{\frac{\alpha }{2p}}\delta _{\mu }^{a}{\rm d}x^{\mu }.  \label{p=alfa}
\end{equation}
Then the singularity (\ref{si}) becomes $e^{a}\simeq {\frac{\rho
_{i}^{\alpha }}{|x-x_{i}|^{\alpha }}}{\rm d}x^{a}$. For $\alpha =1$ the
punctures are of the kind $\relax{\rm I\kern-.18em R}\times S^{3}$  
rather
than $\relax{\rm I\kern-.18em R}^{4}$. It would seem, at first,  
that such
punctures are the most natural candidates to generalize the  
two-dimensional
cylindrical punctures, instead of the ones that we have used so  
far. This,
however, is incorrect. Indeed, if we modify (\ref{p=0}) like  
(\ref{p=alfa}),
we see that a generic $\alpha $ does not produce any change in the  
quantum
topological invariants. In particular, one observes the same kind  
of jumps
as before. Nevertheless, (\ref{p=alfa}) do not share this universality
property. Following the general recipe, one can easily solve the  
topological
field theory with (\ref{p=alfa}). I shall not repeat the derivation  
here,
leaving it to the reader. The normalization (\ref{norm}) is now  
replaced by 
\[
-{\frac{1}{16\pi ^{2}\alpha ^{2}(3-\alpha )}}\int_{\relax{\rm  
I\kern-.18em R}%
^{4}}R^{ab}\wedge R^{cd}\varepsilon _{abcd}=n. 
\]
The technical simplifications occur for $\alpha =p$. Then, the  
punctures of
type $\relax{\rm I\kern-.18em R}\times S^{3}$ correspond to $p=1$. The
characteristic jump (\ref{jump}) can be easily computed to be  
replaced by 
\begin{equation}
<[{\cal O}(x)]^{2}>={\frac{(p^{2}-32p+80)}{40\,(3-p)^{2}}}.   
\label{newjump}
\end{equation}
The punctures of type $\relax{\rm I\kern-.18em R}\times S^{3}$, in
particular, have a jump of $49/160$, a number which is very difficult to
interpret. The right hand side of (\ref{newjump}) has a maximum  
equal to $%
22/35$ for $p=32/13$. We conclude that the jump in not universal in four
dimensions, while it is so in two dimensions. The above expression  
does not
single out anything special about $p=2$.

Let us consider, now, the correlation function $<[\tilde{{\cal  
O}}(x)]^{2}>$%
, where the tilded observables are normalized by replacing the  
$32\pi ^{2}$
with $8\pi ^{2}\alpha ^{2}(3-\alpha )$ in (\ref{tilde}). Then, after a
certain nontrivial amount of work, we get 
\[
<[\tilde{{\cal O}}(x)]^{2}>={\frac{(2-p)(p+10)}{40\,(3-p)^{2}}}, 
\]
which vanishes only for $p=2$ and equals $11/160$ for $p=1$. Thus,  
for $%
p\neq 2$ the quantum mixing between Euler and Pontrijagin numbers  
is much
more apparent than for $p=2$. We finally see that the $\relax{\rm  
I\kern%
-.18em R}^{4}$ punctures are indeed the privileged ones.

We have just learned something very nontrivial from the analysis of the
quantum topological invariants. This shows once again that these  
concepts
are very useful and that the method for treating them developed in \cite
{anomali,me,scond} is very powerful.

\section{Physics}

\label{handles} \setcounter{equation}{0}

In this section I comment about the physical relevance of the  
metrics (\ref
{p=1}) and the notion of punctures in quantum gravity. Moreover, I  
study the
case of a fermion placed in a given quantum topological background.  
For the
issues of gauge invariance and renormalization in connection with the
topological embedding in a general quantum field theory, the reader is
referred to \cite{me,scond}.

The metrics (\ref{p=1}) are instantons of Weyl gravity, but they  
are also
important in ordinary quantum gravity, as we wish to show.

For $p=1$ the Einstein action $\int R\sqrt{g}$ is zero,  
independently of the
number of punctures, since the Ricci curvature vanishes identically.
However, one has to take into account that the total action contains a
boundary term \cite{PAC}, which is not zero for $p=1$. It can be easily
verified that for $p>1$ the boundary term vanishes, while the Einstein
action is negative definite.

According to ref.\ \cite{PAC}, given a metric $g$ one has to look  
for the
element $\bar g$ in the conformal class of $g$ that satisfies  
$R\equiv 0$.
The Positive Action Conjecture \cite{PAC} says that the action for  
$\bar g$,
which is entirely given by the boundary term, is positive definite. The
action for $g$ has an additional contribution, due to the conformal  
factor,
that is positive definite once the integration over the conformal  
factor is
taken to be parallel to the imaginary axis \cite{PAC}. In our case,  
$g$ is a
generic metric (\ref{p=1}), $\bar g$ is the metric with $p=1$. So, the
metrics with $p=1$ are the ones with the least action in our class.

Since the behaviour of $\bar{g}_{\mu \nu }$ for $x\rightarrow  
\infty $ is $%
\delta _{\mu \nu }(1+\sum_{i=1}^{n}\rho _{i}^{2}/x^{2})^{2}$, the  
action for 
$\bar{g}$ is simply \cite{GibPop} 
\begin{equation}
S=-{\frac{1}{16\pi \kappa ^{2}}}\int_{M}{\rm  
d}^{4}x\sqrt{g}R+{\frac{1}{8\pi
\kappa ^{2}}}\int_{\partial M}[K]\sqrt{k}d^{3}x+{\frac{i\theta  
}{64\pi ^{2}}}%
\int_{M}R^{ab}R^{cd}\varepsilon _{abcd}=6\pi \sum_{i=1}^{n}{\frac{\rho
_{i}^{2}}{\kappa ^{2}}}-in\theta .  \label{action}
\end{equation}
Integrating over $\rho _{i}$ gives unity: 
\[
C{\frac{1}{\kappa ^{n}}}\int \prod_{i=1}^{n}{\rm d}\rho  
_{i}\,\,{\rm e}^{-S}=%
{\rm e}^{in\theta }, 
\]
where $C$ is an appropriate numerical factor. This means that the
contribution of the metrics (\ref{p=1}) to the gravitational path  
integral
is, after integrating over all possible values of the scales $\rho  
_{i}$, of
the same order of magnitude as the contribution coming from flat space.
Hence, the metrics (\ref{p=1}) should be relevant for quantum gravity.
Indeed, they can be considered as the analogues of the Yang-Mills  
instantons.

Note that the action (\ref{action}) does not have a minimum in the
topological sector specified by (\ref{norm}): tuning the values of the
scales $\rho _{i}$ appropriately, $S$ can be made arbitrarily  
small. On the
other hand, when some $\rho _{i}$'s are exactly zero, the metric  
belongs to
a different topological sector, since (\ref{norm}) jumps. A  
phenomenon like
this one does not hold for Yang-Mills instantons, where the minimum  
of the
action in the $k^{{\rm th}}$ instanton sector is $8\pi  
^{2}|k|/g^{2}$; it
holds in four dimensional quantum gravity as a consequence of the
peculiarity of the gravitational action (its linearity in the  
curvature) and
of the fact that the boundary of the moduli space of p$\relax{\rm  
I\kern%
-.18emR}_{n}^{4}$ is a union of moduli spaces of p$\relax{\rm  
I\kern-.18emR}%
_{k}^{4}$, $k<n$. In two dimensions, on the other hand, gravity is too
simple to have an effect like this. The practical consequence of this
observation is that in four dimensional quantum gravity,  
differently from
Yang-Mills theory, the topological embedding acts on the positions  
$x_{i}$
of the punctures only, while the scales $\rho _{i}$ are taken care  
of by the
measure $C\prod_{i}{\rm d}\left( \rho _{i}/\kappa \right)  
\hspace{0.03in}%
{\rm e}^{-S}$.

Thus, we understand that the best way to treat the topological  
sectors (\ref
{norm}) is to integrate over all possible values of the $\rho  
_{i}$'s. In
perturbative quantum gravity around the metrics (\ref{p=1}) (with $p=1$)
there will be a linear term in the graviton. It would be interesting to
explore the effects of this linear term in connection with the  
topological
embedding. This, however, is beyond the scope of the present paper  
and maybe
will be considered elsewhere.

The integration over the scales $\rho _{i}$ with the measure  
$C\prod_{i}{\rm %
d}\left( \rho _{i}/\kappa \right) \hspace{0.03in}{\rm e}^{-S}$ does not
affect the topological invariants considered in the past section,  
since they
are scale-independent.

Instead of studying the topological embedding on gravity itself, it is
simpler to start from its effects on matter. Let us consider a fermion
coupled to the gravitational field in $d$ dimensions (we take $p=1$  
if $d=2$
and $p=2$ if $d=4$). The action has the additional contribution 
\[
\sqrt{g}\overline{\psi }\gamma ^{a}e_{a}^{\mu }{\cal D}_{\mu }\psi  
={\rm e}^{%
{\frac{d-1}{2}}\varphi }\overline{\psi }{\cal D}\!\!\!\!\slash \psi . 
\]
The presence of the factor ${\rm e}^{\varphi }$, which is singular  
in the
puctures, forces the matter field to vanish there, more or less as it
happens in the case of vortices \cite{scond}.

We are interested in the two-point function $<\psi  
(x)\overline{\psi }(y)>$
in a punctured background. This means that we have to integrate over the
positions of the puncures. This integral is, in general, divergent.  
However,
according to the general theory developed in \cite{me} and \cite{scond},
certain amplitudes are order-by-order perturbatively well-defined in the
topologically nontrivial sectors. Here we want to apply those ideas  
to our
case. One considers, instead of the ill-defined $<\psi  
(x)\overline{\psi }%
(y)>$, amplitudes of the form 
\begin{equation}
<\psi (x)\overline{\psi }(y)\prod_{j=1}^{n}{\cal O}(z_{j})>\equiv  
<\psi (x)%
\overline{\psi }(y)>_{\{z\}}.  \label{pop1}
\end{equation}
The insertion of the local observables ${\cal O}(z_{j})$ ($n$ of  
them, in
the case of $n$ punctures) specifies the quantum topological  
background on
which the amplitude is defined. Intuitively, it is like studying phonon
interactions on the background of the magnetic force lines that  
penetrate a
type II superconductor \cite{scond}. An amplitude like the above  
one will be
called a $(2,n)$-point function, the first number referring to the  
quantum
excitations, the second number referring to the quantum background.  
Later on
we shall also compute the fermion two-point function with a  
nonlocal quantum
background made of curves $\gamma $.

Before beginning the computations, we have to make a couple of comments
about the solutions (\ref{unmod}) and (\ref{unmod2}) and the  
meaning of the
topological gauge-fixing conditions on the physical side of the problem.

It has been shown in \cite{anomali} that the gauge-fixing condition  
for the
topological ghosts uniquely determine the solution to the  
topological field
theory. In our case, $\psi ^{a}=\psi ^{ab}e^{b}$ should be fixed  
with the
conditions $\psi ^{ab}=\psi ^{ba}$ and ${\cal D}^{\mu }\psi _{\mu  
}^{a}=0$.
However, (\ref{unmod}) and (\ref{unmod2}) do not satisfy ${\cal D}^{\mu
}\psi _{\mu }^{a}=0$: (\ref{unmod}) and (\ref{unmod2}) are only smooth
deformations of the ``good'' solutions. The topological invariants are
unaffected and, indeed, in two dimensions we were able to reproduce some
known results. In the analysis of the physical side of the problem,  
on the
other hand, the gauge-fixing condition for $\psi ^{a}$ is  
important. One can
check that the physical amplitudes, like (\ref{pop1}), do depend on  
it. As
far as we know, it could be impossible to write down a gauge  
condition for $%
\psi ^{a}$ that rigorously produces the solutions (\ref{unmod}) or (\ref
{unmod2}) and satisfies the standard requirements of quantum field  
theory
(locality, in particular). ${\cal D}^{\mu }\psi _{\mu }^{a}=0$  
appears to be
the natural (and, to some extent, unique) choice. On the other hand, the
explicit solution to this equation, that we have not found, could be so
complicated to prevent us from using it for practical purposes.  
Therefore,
we use (\ref{unmod}) and (\ref{unmod2}) in this paragraph\footnotemark 
\footnotetext{
The explicit computations will be mainly performed in two  
dimensions, since
this is already a nontrivial and illustrative case, but simpler.},  
taking
full advantage of their simplicity, but being aware that in any  
hypothetical
``comparison with experiment'' our results should be mostly taken as
qualitative.

In full generality, the gauge-fixing conditions ${\cal G}_{\psi  
}=0$ for the
topological ghosts $\psi _{\mu }^{a}$ appear to be as important as the
instanton conditions ${\cal G}_{T}=0$ themselves (for example, in  
Yang-Mills
theory, ${\cal G}_{T}=F_{{}}^{a+}$ and ${\cal G}_{\psi }=D^{\mu  
}\psi _{\mu
}^{a}$). This is quite reasonable, since the topological ghosts  
$\psi _{\mu
}^{a}$ are descendants of the curvature ($F^{a}$, $R^{ab}$, $\ldots  
$). It
is for this reason, for example, that the non-covariant choice  
${\cal G}%
_{\psi }=\partial ^{\mu }\psi _{\mu }^{a}$ is not acceptable:  
indeed, in 
\cite{anomali} it was observed that ${\cal G}_{\psi }=\partial  
^{\mu }\psi
_{\mu }^{a}$ leads to an empty topological Yang-Mills theory. The  
instanton
conditions ${\cal G}_{T}=0$ are also a sort of gauge-fixing  
conditions, from
the topological field theoretical viewpoint, but on the physical  
side they
play quite a different role, because they are uniquely determined by
minimizing the action.

Summarizing, changing ${\cal G}_{\psi }$ does change the physical
amplitudes, but, fortunately, there exists a natural choice for it.
Moreover, ${\cal G}_{\psi }$ is strictly unique if one accepts that it
should combine together with the $\psi $-field equation into a  
twisted Dirac
equation. This ``gauge-fixing dependence'' is not in contraddiction with
general principles. The point is that we are talking about the  
gauge-fixing
conditions of the topological symmetry, which is {\sl not }a  
symmetry of the
physical theory, but rather an artifact of the perturbative  
expansion around
instantons. The topological ``gauge-symmetry'' is present because in the
topologically nontrivial sectors the minimum of the action is not a  
point,
but a moduli space. We can say that the topological embedding solves the
problem that arises when the very first stage of the perturbative  
expansion
(with which I mean the zero point function) has more  
``gauge-symmetries''
than the complete theory. In the topological embedding the quantum
topological invariants play the very role of ``zero point functions''.

After this instructive digression, let us begin our analysis of the  
physical
amplitudes.

In two dimensions the action of pure gravity is topological. The  
scales $%
\rho_i$ deserve to be discussed apart. In this respect, the situation is
different from the four dimensional one. The $\rho_i$'s can or cannot be
integrated over, according to the background that one wish to consider.
Here, I shall examine both cases, starting from the case in which the $%
\rho_i $ are fixed. The amplitudes can be $\rho_i$-dependent.

The covariant derivative ${\cal D}\psi $ is ${\rm d}\psi -{\frac{i}{2}}%
\omega \sigma _{3}\psi $. Defining $\overline{\psi }={\rm e}^{-{\frac{%
\varphi }{4}}}\widetilde{\overline{\psi }}$ and $\psi ={\rm  
e}^{-{\frac{1}{4}%
}\varphi }\widetilde{\psi }$, the lagrangian can be written simply as 
\[
{\rm e}^{\frac{\varphi }{2}}\overline{\psi }\sigma ^{\mu }\left(  
\partial
_{\mu }+{\frac{1}{4}}\partial _{\mu }\varphi \right) \psi =\widetilde{%
\overline{\psi }}\sigma ^{\mu }\partial _{\mu }\tilde{\psi}, 
\]
where ${\rm e}^{\varphi (x)}=D(x)=1+\sum_{i=1}^{n}{\frac{\rho  
_{i}^{2}}{%
(x-x_{i})^{2}}}$. Thus, the desired $(2,n)$-point function is  
easily written
down. One has 
\begin{equation}
<\psi (x)\overline{\psi }(y)>_{\{z\}}={\frac{1}{2\pi  
}}{\frac{x\!\!\!\slash%
-y\!\!\!\slash }{(x-y)^{2}}}G_{n}(x,y,\{z\}),  \label{rut}
\end{equation}
where the function $G(x,y,\{z\})$ is a sort of form factor that  
describes
the deviation from the free propagator. In order to simplify the  
expression,
I would like to consider the case in which the $n$ points $z_{i}$  
are all
distinct (coincidences appear to be quite unplausible, from the physical
point of view), but concentrated in a small region (or viewed from  
a large
distance). In this way, we can take $z_{j}\simeq z$ $\forall j$.  
Were the
punctures truly placed in the same point $z$, there would be a jump
described by an additional overall factor $1/n!$, as we know.

Under these conditions, we have, from (\ref{urc}), 
\begin{equation}
G_{n}(x-z,y-z)={\frac{n!}{\pi ^{n}}}\int \prod_{i=1}^{n}{\frac{\rho  
_{i}^{2}%
{\rm  
d}^{2}x_{i}}{(z-x_{i})^{4}}}\,\,{\frac{1}{D(z)^{n+1}D(x)^{\frac{1}{4}
}D(y)^{\frac{1}{4}}}}  \label{fgfg}
\end{equation}
Although it was not obvious from the beginning, the above formula  
shows that
the amplitudes that we are considering are indeed well-defined: the  
$x_{i}$%
-integrations make sense. This would not be true without the  
insertion of
the quantum background $\prod_{j=1}^{n}{\cal O}(z_{j})$. Thus we have a
concrete illustration and check of the general theory developed in  
ref.s 
\cite{me,scond}. (\ref{rut}) represents the first perturbative  
contribution
to the effective action in the $n$-puncture topological sector,  
obtained by
expanding around the topological amplitude $<\prod_{j=1}^{n}{\cal O}%
(z_{j})>=1$\footnotemark 
\footnotetext{
The words ``expanding around a topological amplitude'' should not  
suggest
that there is a smooth limit that reduces the physical theory to the
topological one. Indeed, it is true that the quantum fluctuations are
``small'', but it is also true that, after the functional  
integration, their
contribution is finite.}. The result is no more topological, of  
course, and
the factor $G(x-z,y-z)$ measures the feedback that the fermion  
propagation
receives from the quantum background on which it is excited. $G$  
could be
considered as an effective ``quantum'' metric. In our examples $G$  
obeys the
inequality $0\leq G\leq 1$. The amount of the deviation of $G$ from the
value 1 (propagator in flat space) describes the ``obstacle'' that the
fermion finds on its way.

This phenomenon, i.e.\ the generation of a quantum metric by the quantum
background, seems to be quite general, not related to the presence of
gravity in the problem. For example, a two-point function of the same
structure can be observed (see formula (2.8) of \cite{scond}) with  
scalars
in presence of the BPST instanton \cite{belavin}.

In our simple theory, there is no radiative correction to the effective
action and the form factor $G(x-z,y-z)$ is the entire story.

Let us focus on the singular behaviour for $x\sim y$. The function $%
G_{n}(x-z,x-z)$ has a minimum exactly on the puncture ($x=z$) and  
tends to 1
far from it: the deviation from the free propagator is maximal  
nearby the
puncture and negligible elsewhere. The minimum $G_{n}(0,0)$ is  
$\frac{(2n)!!%
}{(2n+1)!!}$, so that 
\[
<\psi (x)\overline{\psi }(z)>_{z}\sim  
{\frac{(2n)!!}{(2n+1)!!}}{\frac{1}{%
2\pi }}{\frac{x\!\!\!\slash-z\!\!\!\slash }{(x-z)^{2}}}. 
\]
For large $n$ this minimum goes to zero like ${\frac{1}{n}}$. This means
that an infinite number of punctures is able to inhibit the propagation
completely in the area where they are located. This is a quite  
reasonable
physical result. But it is also interesting to note that no finite  
number of
punctures is able to achieve this.

We can compare the effects of the quantum background to the ones of the
classical background. The latter situation is achieved by saying  
that the
metric is a fixed external field, so that no integrations over the  
positions 
$x_{i}$ are performed. Then, the fermion two-point function 
\begin{equation}
<\psi (x)\overline{\psi }(y)>={\frac{1}{2\pi }}{\rm  
e}^{-{\frac{\varphi (x)}{%
4}}}{\frac{x\!\!\!\slash-y\!\!\!\slash}{(x-y)^{2}}}\,{\rm e}^{-{\frac{%
\varphi (y)}{4}}}  \label{clmetric}
\end{equation}
vanishes at the points $x_{i}$, whatever the total number of  
puctures is. We
conclude that the quantum background is able to smoothen this effect.

For one puncture the form factor is 
\begin{equation}
G(x-z,y-z)={\frac{1}{\pi }}\int {\frac{\rho ^{2}{\rm d}^{2}x_{0}}{(\rho
^{2}+(z-x_{0})^{2})^{2}}\frac{1}{\left[ \left( 1+{\frac{\rho ^{2}}{%
(x-x_{0})^{2}}}\right) \left( 1+{\frac{\rho ^{2}}{(y-x_{0})^{2}}}\right)
\right] ^{\frac{1}{4}}}}  \label{hk}
\end{equation}
and its minimum is $\frac{2}{3}$. We understand that the physical  
meaning of
the scale $\rho $ is that it measures the size of the region where the
propagation is sensibly affected by the $G$-function.

Now, we want to analyse a different possibility for defining $<\psi (x)%
\overline{\psi }(y)>$ in the one-puncture sector. We want to insert  
a couple
of nonlocal observables like the ones appearing in (\ref{2.11}). We  
have 
\[
<\psi (x)\overline{\psi }(y)>_{\gamma _{1}\cdot \gamma  
_{2}}={\frac{1}{2\pi }%
}{\frac{x\!\!\!\slash-y\!\!\!\slash }{(x-y)^{2}}}G(x,y,\gamma _{1}\cdot
\gamma _{2}). 
\]
where 
\[
G(x,y,\gamma _{1}\cdot \gamma _{2})={\frac{\rho ^{4}}{\pi  
^{2}}}\int_{\gamma
_{1}}\int_{\gamma _{2}}\int {\frac{\varepsilon _{\mu \nu }{\rm  
d}z^{\mu }%
{\rm d}w^{\nu }\,\,{\rm d}^{2}x_{0}}{(\rho ^{2}+(z-x_{0})^{2})^{2}(\rho
^{2}+(w-x_{0})^{2})^{2}\left[ \left( 1+{\frac{\rho  
^{2}}{(x-x_{0})^{2}}}%
\right) \left( 1+{\frac{\rho ^{2}}{(y-x_{0})^{2}}}\right) \right]  
^{\frac{1}{%
4}}}}. 
\]
To be explicit, let us $\gamma _{1}$ be the $x$-axis and $\gamma  
_{2}$ the $%
y $-axis. The $\gamma $-integrations are easily doable. The minimum  
of $G$
is in the point in which $\gamma _{1}$ and $\gamma _{2}$ intersect. A
numerical integration gives, for $x\sim y\sim 0$ 
\[
<\psi (x)\overline{\psi }(y)>_{\gamma _{1}\cdot \gamma _{2}}\sim  
0.6935\,{%
\frac{1}{2\pi }}{\frac{x\!\!\!\slash-y\!\!\!\slash }{(x-y)^{2}}}, 
\]
revealing that the deviation from free propagation due to two  
intersecting
lines is slightly less than the one due to a local observable. Note,
however, that the $\gamma $'s distribute $G$ in a wider area.

Finally, let us study the topological embedding when the scales  
$\rho _{i}$
are included in the set of the moduli. We study the propagation on the
quantum background (\ref{quantumbackground}). We have 
\begin{equation}
<\psi (x)\overline{\psi }(y)>_{\gamma \cdot \{z\}}={\frac{1}{2\pi  
}}{\frac{%
x\!\!\!\slash-y\!\!\!\slash }{(x-y)^{2}}}G(x,y,\gamma \cdot \{z\}).
\label{pop2}
\end{equation}
where 
\begin{equation}
G(x,y,\gamma \cdot \{z\})={\frac{1}{4\pi ^{2}}}\int_{\gamma }\int  
{\frac{%
{\rm d}w^{\mu }\varepsilon _{\mu \nu }(w-z)^{\nu }\,2\rho ^{2}{\rm  
d}\rho
^{2}\,{\rm d}^{2}x_{0}}{(\rho ^{2}+(z-x_{0})^{2})^{2}(\rho
^{2}+(w-x_{0})^{2})^{2}\left[ \left( 1+{\frac{\rho  
^{2}}{(x-x_{0})^{2}}}%
\right) \left( 1+{\frac{\rho ^{2}}{(y-x_{0})^{2}}}\right) \right]  
^{\frac{1}{%
4}}}}.  \label{oppo}
\end{equation}
The integral in $\rho $ and $x_{0}$ diverges as  
$\frac{1}{(z-w)^{2}}$ when $%
w\rightarrow z$, precisely as in the purely topological amplitude (\ref
{quantumbackground}). The factor in front of the singularity,  
however, can
be different. An instructive situation that we can easily handle is  
the one
in which $x\sim y\sim z$. In other words, let us inspect how the  
propagator
is modified in the neighborhood of the point $z$ where the local  
observable
is placed. One gets 
\begin{equation}
<\psi (x)\overline{\psi }(y)>_{\gamma \cdot \{x\}}\sim {\frac{2}{3}}%
\,\backslash \!\!\!\slash (\gamma ,\{x\})\,{\frac{1}{2\pi  
}}{\frac{x\!\!\!%
\slash-y\!\!\!\slash }{(x-y)^{2}}}.  \label{4.16}
\end{equation}
We see that there are both the numerical factor $2/3$ and the link  
number
itself in front of the usual two-point function. At very large  
distances, on
the other hand, the two point function looses the factor $2/3$ and  
becomes
the free propagator times the link number. An intermediate  
situation is the
one in which one point, say $x$, is very far and the other one is  
very close
to $z$. Then the factor $2/3$ in (\ref{4.16}) is replaced by $4/5$.  
If the
curve $\gamma $ and the point $z$ are unlinked, (\ref{4.16}) is exactly
zero, however the two-point function is not identically zero.

Since $\rho$ is integrated over, there is no external scale which  
is kept
fixed. The size of the region where the propagation is sensibly  
affected by
the quantum background is dictated by the quantum background itself:
precisely it is the size of the curve $\gamma$.

We have examined two typical cases. In the first case the  
instantons contain
a dimensionful parameter $\rho$ that is kept fixed. Then the quantum
background can be constructed purely with local observables, like  
in (\ref
{pop1}). In the second case, instead, the size of the instanton is
integrated over. Then the quantum background is forced to contain  
at least
one nonlocal observable, like in (\ref{pop2}), associated with a closed
submanifold $\gamma$. The brings a new scale into the game, which is the
size of $\gamma$. It seems that it is not possible to have a scale-free
topological embedding.

Let us recall that, in the topologically trivial sector, when no  
other scale
is around, it is renormalization that forces the introduction of one. In
section 2.1 of ref. \cite{scond} the usual renormalization scale was
interpreted as the (unique) quantum background of the topologically  
trivial
sector. Indeed, a feature that the topological embedding and  
renormalization
have in common is that they both cure problems with divergences.

It seems that the factor $G$ has everywhere the same sign, which we  
can fix
conventionally to be positive. I have not found situations where this
property is violated. It is straightforward to prove it for the  
amplitude (%
\ref{fgfg}) and for a set of simple situations in (\ref{oppo}) (for  
example,
when $\gamma $ is a circle centered in $z$). However, I do not have a
rigorous proof that this positivity condition holds in general. It would
assure that our two-point function is physically meaningful without  
imposing
restrictions on the quantum background. At the same time, it would  
justify
the name ``quantum metric'' that we have used.

The analysis of the two-point function of a Dirac fermion in a four
dimensional punctured background is entirely similar and will not be
repeated. We just note that it is sufficient to write the  
Lagrangian $L=%
\sqrt{g}\overline{\psi }\gamma ^{a}e_{a}^{\mu }{\cal D}_{\mu }\psi  
$ as $L=%
\widetilde{\overline{\psi }}\partial \!\!\!\slash\widetilde{\psi }$  
for $%
\widetilde{\overline{\psi }}=\overline{\psi }\hspace{0.03in}{\rm  
e}^{\frac{3%
}{2p}{\varphi }}$ and $\widetilde{\psi }=\psi \hspace{0.03in}{\rm  
e}^{\frac{3%
}{2p}{\varphi }}$ (${\cal D}\psi ={\rm d}\psi -\frac{1}{8}[\gamma
_{a},\gamma _{b}]\omega ^{ab}\psi $ in our notation) and proceed as  
before,
using the moduli-space measures of section \ref{punct4}. Similarly,  
messless
QED (or QCD, with obvious modifications) in the same punctured  
background
presents no further difficulty. The QED\ Lagrangian in the variables $%
\widetilde{\overline{\psi }}$ and $\widetilde{\psi }$ still looks  
like the
ordinary one. Consequently, denoting by $\Gamma \left(  
x,m;y,m;z,l\right) $
the ordinary renormalized QED scattering amplitude of $m$ fermions  
and $l$
photons, the same amplitude in the quantum punctured background $Q$  
is given
by $\Gamma _{Q}\left( x,m;y,m;z,l\right) =\Gamma \left(  
x,m;y,m;z,l\right)
G_{Q}\left( x,m;y,m\right) $, where ($p=1$)

\begin{equation}
G_{Q}\left( x,m;y,m\right) =C\int \prod_{i=1}^{n}{\rm d}\left( \rho
_{i}/\kappa \right) \hspace{0.03in}{\rm  
e}^{-S_{Q}}\hspace{0.03in}{\rm d\mu }%
_{Q}\prod_{j=1}^{m}{\rm e}^{-\frac{3}{2}\left( {\varphi (x}_{j})+\varphi
(y_{j})\right) }.  \label{iss}
\end{equation}
Here ${\rm d\mu }_{Q}$ denotes the measure of the topological  
amplitude $%
\int {\rm d\mu }_{Q}$ associated with the quantum background $Q$.  
$S_{Q}$ is
the action (\ref{action}) of the background $Q$. It is  
straightforward to
check that $G_{Q}$ is well-defined and that its qualitative  
behaviour agrees
with the behaviours of the $G$-functions studied so far. A direct
consequence of (\ref{iss}) is that the pure scattering of photons  
($m=0$)
does not really feel the quantum background, in the sense that  
$G_{Q}$ is
just the topological invariant in that case. (\ref{iss}) could have some
physical applications, for example in solid state physics, if one  
finds some
material that is able to simulate the effects of the punctures. In this
hypothetical situation, (\ref{iss}) could describe the QED  
scattering inside
such a material. Finally, we can see explicitly that there is no  
conflict
between the topological embedding and renormalization.

In conclusion, we have tested the general theory of the topological
embedding in a set of simple models in which we can write down explicit
formul\ae . The amplitudes are well-defined and give quite plausible
physical predictions. We have seen that the known properties of quantum
field theory are generalized in a reasonable way. Up to now, the  
analysis of
the topological embedding has not revealed the need of physical  
restrictions
on the quantum background one is expanding around. The topological  
embedding
could be a new chapter of quantum field theory.


\begin{thebibliography}{99}
\bibitem{scond}  D.\ Anselmi, On field theory quantization around
instantons, preprint HUTP-95/A026 and hepth/9507167, July 1995.

\bibitem{anomali}  D.\ Anselmi, Anomalies in instanton calculus, Nucl.\
Phys.\ B439 (1995) 617.

\bibitem{me}  D.\ Anselmi, Topological field theory and physics,  
preprint
HUTP-95/A012 and hepth/9504049, March 1995.

\bibitem{thooft}  G.\ 't~Hooft, Computation of the quantum effects  
due to a
four-dimensional pseudoparticle, Phys.\ Rev.\ D 14 (1976) 3432.

\bibitem{twist1}  D.\ Anselmi and P.\ Fr\'{e}, Twisted N=2  
supergravity as
topological gravity in four dimensions, Nucl.\ Phys.\ B392 (1993) 401.

\bibitem{kontsevich}  M.\ Kontsevich, Intersection theory on the moduli
space of curves and the matrix Airy function, Comm.\ Math.\ Phys.\ 147
(1992) 1.

\bibitem{constr}  D.\ Anselmi, P.\ Fr\'{e}, L.\ Girardello and P.\  
Soriani,
Constrained topological field theory, Phys.\ Lett.\ B 335 (1994) 416.

\bibitem{GibPop}  G.W.\ Gibbons and C.N.\ Pope, The Positive Action
Conjecture and asymptotically Euclidean metrics in quantum gravity,  
Cummun.\
Math.\ Phys.\ 66, 267 (1979).

\bibitem{rebbi}  R.\ Jackiw, C.\ Nohl and C.\ Rebbi, Conformal  
properties of
pseudoparticle configurations, Phys.\ Rev.\ D 15 (1977) 1642.

\bibitem{PAC}  G.W.\ Gibbons, S.W.\ Hawking and M.J.\ Perry, Path  
integral
and the indefiniteness of the gravitational action, Nucl.\ Phys.\ B138
(1978) 141.

\bibitem{belavin}  A.A.\ Belavin, A.M.\ Polyakov, A.S.\ Schwarz and  
Yu.S.\
Tyupkin, Pseudoparticle solutions of the Yang-Mills equations,  
Phys.\ Lett.\
59B (1975) 85.
\end{thebibliography}
\end{document}